\documentclass[aps,prl,groupedaddress]{revtex4}

\usepackage{graphicx}

\begin{document}



\title{The A-B Transition of Superfluid $^3$He in Aerogel and the Effect of Anisotropic Scattering}

\author{C.L. Vicente$^1$, H.C. Choi$^2$, J.S. Xia$^1$, W.P. Halperin$^3$, N. Mulders$^4$, and Y. Lee$^{1,2}$}
\email[]{yoonslee@phys.ufl.edu} \affiliation{ $^1$National High
Magnetic Field Laboratory, Gainesville, FL
32611-8440\\$^2$Department of Physics, University of Florida,
Gainesville FL 32611-8440\\$^3$Department of Physics and
Astronomy, Northwestern University, Evanston IL
60208\\$^4$Department of Physics, University of Delaware, Newark
DE 19716}


\date{\today}

\begin{abstract}
We report the results of high frequency acoustic shear impedance
measurements on superfluid $^3$He confined in 98\% porosity silica
aerogel. Using 8.69 MHz continuous wave excitation, we measured
the acoustic shear impedance as a function of temperature for the
sample pressures of 28.4 and 33.5 bar. We observed the A-B
transition on warming in zero magnetic field. Our observations
show that the A and B phases in aerogel coexist in a temperature
range of about 100 $\mu$K in width.  We propose that the relative
stability of the A and B phases from anisotropic scattering can
account for our observations.
\end{abstract}

\maketitle


The effect of disorder on a condensed matter system is one of the
most interesting and ubiquitous problems in condensed matter
physics.  Metal-insulator transitions \cite{paLee} and the Kondo
effect \cite{heeger} are two examples of phenomena in which
disorder, in the form of various types of impurities, plays a
fundamental role. The influence of disorder on ordered states such
as the magnetic or superconducting phase has also attracted
tremendous interest especially in systems that undergo quantum
phase transitions \cite{sachdev}. In this work we address the
influence of disorder on the first order phase transition that
separates two highly competing ordered states. This situation
occurs in superfluid $^3$He impregnated in high porosity silica
aerogel where the ordered states are the A and B phases with {\it
p-wave} spin-triplet pairing \cite{vollhardt}.

High porosity silica aerogel consists of a nano-scale abridged
network of SiO$_2$ strands with a diameter of 3 $\sim$ 5~nm. Since
the diameter of the strands is much smaller than the coherence
length ($\xi_o$) of the superfluid (72~nm - 15~nm in the pressure
range of 0 to 34 bar), when $^3$He is introduced, the aerogel
behaves as an impurity with the strands acting as effective
scattering centers. The phase diagram of $^3$He/aerogel (with
mostly 98\% porosity) has been studied using a variety of
techniques
\cite{porto,sprague1,mat97,gervais,brussard,baumgardner2,choi}. To
date, three distinct superfluid phases have been observed in
$^3$He/aerogel; these are called the A, B, and A$_1$ phases
\cite{choi} as in the bulk, although the detailed structures of
the order parameters (especially the orbital component) have not
been identified. Early studies using NMR \cite{sprague1} and
torsional oscillator \cite{porto} measurements on $^3$He in 98\%
porosity aerogel found substantial depression in the superfluid
transition and a theoretical account based on the homogeneous
isotropic scattering model (HISM) was provided \cite{thuneberg}.
The NMR studies of $^3$He/aerogel show evidence of an equal spin
pairing (ESP) state similar to the bulk A phase \cite{sprague1}
and a phase transition to a non-ESP state similar to the bulk B
phase \cite{barker}. A large degree of supercooling was observed
in this phase transition (A-B transition) indicating the
transition is first order. Further studies using acoustic
techniques \cite{gervais}, and an oscillating aerogel disc
\cite{brussard}, have confirmed the presence of the A-B transition
in the presence of low magnetic fields.

While the effects of impurity scattering on the second order
superfluid transition have been elucidated by these early studies,
experiments designed to determine the effects of disorder on the
A-B transition have been rather inconclusive
\cite{barker,gervais,brussard,dmitriev,nazaretski,baumgardner}. It
is important to emphasize that the free energy difference between
the A and B phases in bulk $^3$He is minute compared to the
condensation energy \cite{leggett}. Moreover, both phases have
identical intrinsic superfluid transition temperatures. The nature
of highly competing phases separated by the first order transition
is at the heart of many intriguing phenomena such as the
nucleation of the B-phase in the meta-stable A-phase environment
\cite{leggett}, the profound effect of magnetic fields on the A-B
transition \cite{paulson}, and the subtle modification of the A-B
transition in restricted geometry \cite{li}. We expect this
transition to be extremely sensitive to the presence of aerogel
and conjecture that even the low energy scale variation of the
aerogel structure would have a significant influence on the A-B
transition.

A number of experiments have been performed with the purpose of
systematically investigating the A-B transition in aerogel
\cite{barker,gervais,brussard,nazaretski,baumgardner}. In
experiments by the Northwestern group \cite{gervais} using a shear
acoustic impedance technique, a significantly supercooled A-B
transition was seen while no signature of the A-B transition on
warming was identified. In the presence of magnetic fields,
however, the equilibrium A-B transitions were observed and the
field dependence of the transition was found to be quadratic as in
the bulk. However, no divergence in the coefficient of the
quadratic term, $g(\beta)$ ($1 - T_{AB}/T_c = g(\beta)(H/H_c)^2$,
where $T_{AB}$ is the equilibrium A-B transition temperature) was
observed below the melting pressure. This result is in marked
contrast with the bulk behavior which shows a strong divergence at
the polycritical point (PCP) \cite{hahn}. These authors conclude
that the strong-coupling effect is significantly reduced due to
the impurity scattering and the PCP is absent in this system.
Although this conclusion seems to contradict their observation of
a supercooled A-B transition even at zero field, other theoretical
and experimental estimations of the strong-coupling effect in the
same porosity aerogel \cite{baramidze,choi} seem to support their
result. The Cornell group \cite{nazaretski} investigated the A-B
transition in 98\% aerogel using the slow sound mode in the
absence of a magnetic field.  While the evidence of the
supercooled A-B transition was evident, no warming A-B transition
was observed.  Nonetheless, they observed a partial conversion
from B$\rightarrow$A phase only when the sample was warmed into
the narrow band ($\approx$ 25 $\mu$K) of aerogel superfluid
transition. Recently, the Stanford group conducted low field (H =
284 G) NMR measurements on 99.3\% aerogel at 34 bar
\cite{baumgardner}. They found about 180 $\mu$K window below the
superfluid transition where the A and B phases coexist on warming
with a gradually increasing contribution of the A-phase in the NMR
spectrum.

We have observed the A-B transition on warming in the absence of
magnetic field for two sample pressures of 28.4 and 33.5 bar, and
have found evidence that the two phases coexist in a temperature
window that can be as wide as 100 $\mu$K. In this Letter, we
present our findings along with some considerations that can
clarify the seemingly paradoxical observation on the A-B
transition in aerogel and, at the same time, provide a plausible
explanation for the observed coexistence of the A and B phases.

A continuous wave shear impedance technique was employed in this
study.   In this method, an AC-cut quartz transducer, which is in
contact with bulk (clean) liquid and (dirty) liquid in aerogel, is
excited continuously at a frequency of 8.69~MHz.  Using a bridge
type cw spectrometer, the change in electrical impedance of the
transducer is measured while the temperature was varied slowly.
Typical warming (cooling) rates used in our study are between 0.1
and 0.2~mK/hr. A $^3$He melting pressure thermometer is used as
our main thermometer.  The strength of stray field at the sample
cell from the demagnetization magnet is smaller than 10 G in the
experimental condition.  The experimental technique and the sample
cell are described elsewhere in detail \cite{gervais, choi,
lee99}.

\begin{figure}
\includegraphics[height = 2.5 in, angle = -90]{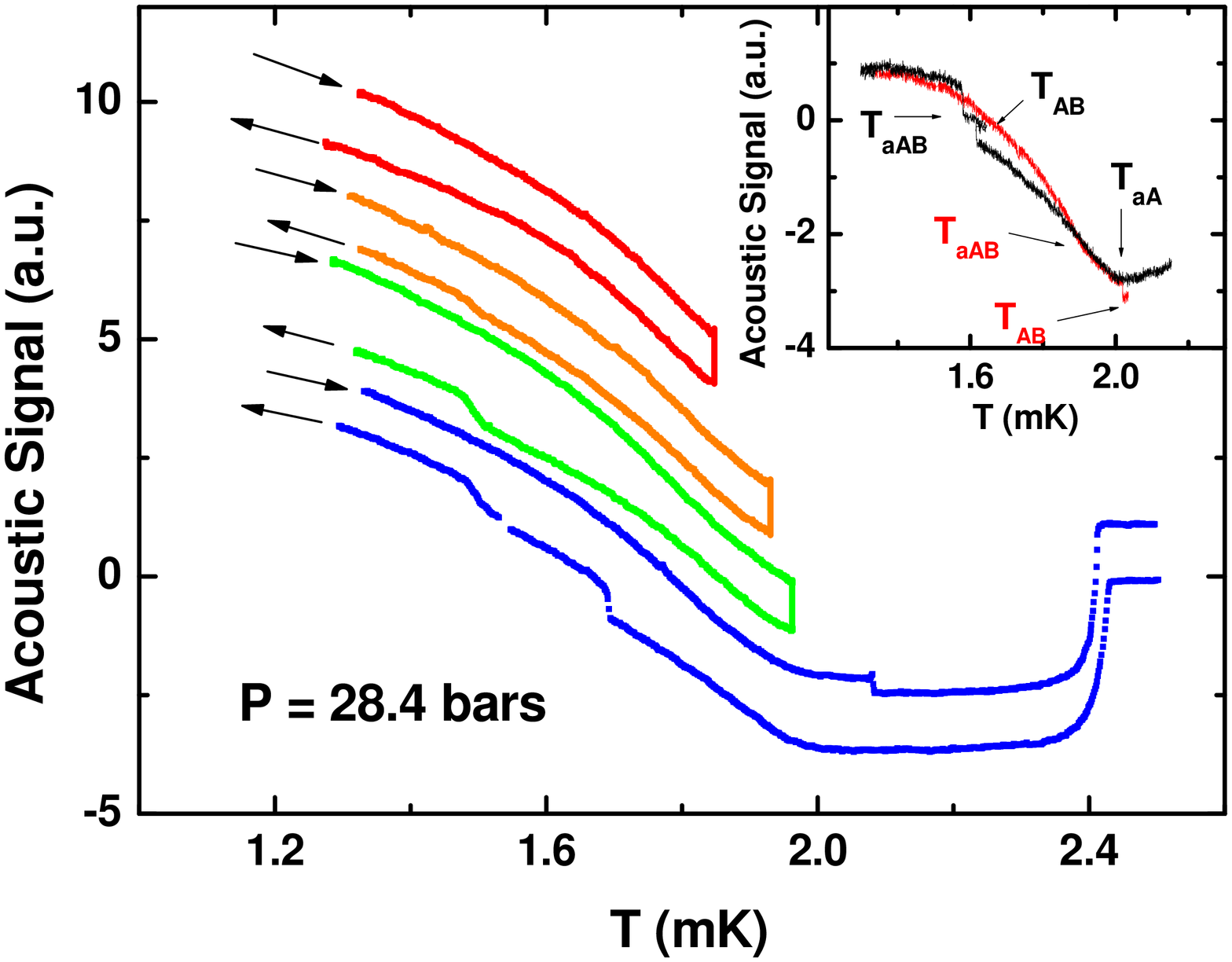}
\caption{\label{track28} Acoustic traces of tracking experiments
at 28.4 bar in zero magnetic field.  Each pair of warming and
subsequent cooling is color-coded. The turn-around temperatures
are indicated by the vertical line for each pair. The arrows
indicate the direction of temperature change in time.  Inset:
cooling (black) and warming (red) traces taken at around 28.5 bar.
The signatures of the aerogel superfluid transition and the
aerogel A-B transition are labelled as $T_{aA}$ and $T_{aAB}$.}
\end{figure}

Shown in Fig.~1 are the traces of acoustic signal taken at 28.4
bar in zero magnetic field.   The bottom (blue) traces show the
acoustic responses between 1.3~mK and 2.5~mK.  The sharp jumps in
the acoustic traces around 2.4~mK mark the bulk superfluid
transition and distinct slope changes are associated with the
superfluid transitions in aerogel (see Ref.~\cite{choi}). The
signatures of the supercooled A-B transition in the bulk and
aerogel appear as small steps on the cooling traces. The
identification of the step in the acoustic impedance as the A-B
transition has been established by a systematic experimental
investigation of Gervais {\it et al.}~\cite{gervais}. In the
inset, the warming and cooling traces near 28.5~bar are shown in a
narrower temperature range near the aerogel superfluid transition.
This trace was the first indication of a possible A-B transition
in aerogel on warming at zero field. The cooling trace (black)
from the normal state of bulk reveals a well-defined aerogel
transition at 2.0~mK ($T_{aA}$).  The supercooled A-B transitions
from the bulk ($T_{AB}$) and aerogel ($T_{aAB}$) are clearly shown
as consecutive steps at lower temperatures. After being cooled
through both A-B transitions, both clean and dirty liquids are in
the B-phase. On warming  the trace follows the B-phase and
progressively merges into the A-phase (cooling trace) around
1.9~mK. This subtle change in slope is the signature of the A-B
transition on warming.

In order to test our identification of this feature in aerogel, we
performed tracking experiments similar to those described in
Ref.~\cite{gervais}. The sample liquid is slowly warmed from the
aerogel B phase up to various points around the feature, and then
cooled slowly to watch the acoustic trace for the signature of the
supercooled A-B transition.  During the turn-around, the sample
stays within 30 $\mu$K from the highest temperature reached
(hereafter referred to as the turn-around temperature) for about
an hour. If the warming feature is indeed the A-B transition,
there should be a supercooled signature on cooling only after
warming through this feature. The color coded pairs of the traces
in Fig.~1 are the typical results of the tracking experiments for
different turn-around temperatures. In the bottom (blue) cooling
trace from the normal state, one can clearly see two supercooled
A-B transition steps. The sharper step appearing at $\approx$
1.7~mK corresponds to the bulk A-B transition. We find that the
size of the step indicating the aerogel A-B transition depends on
the turn-around temperature.  We can make a direct comparison of
each step size since the supercooled A-B transitions in aerogel
occur within a very narrow temperature range, $\approx$ 40~$\mu$K.
Similar behavior was observed for 33.5~bar.

\begin{figure}
\includegraphics[height = 3.0 in, angle = -90]{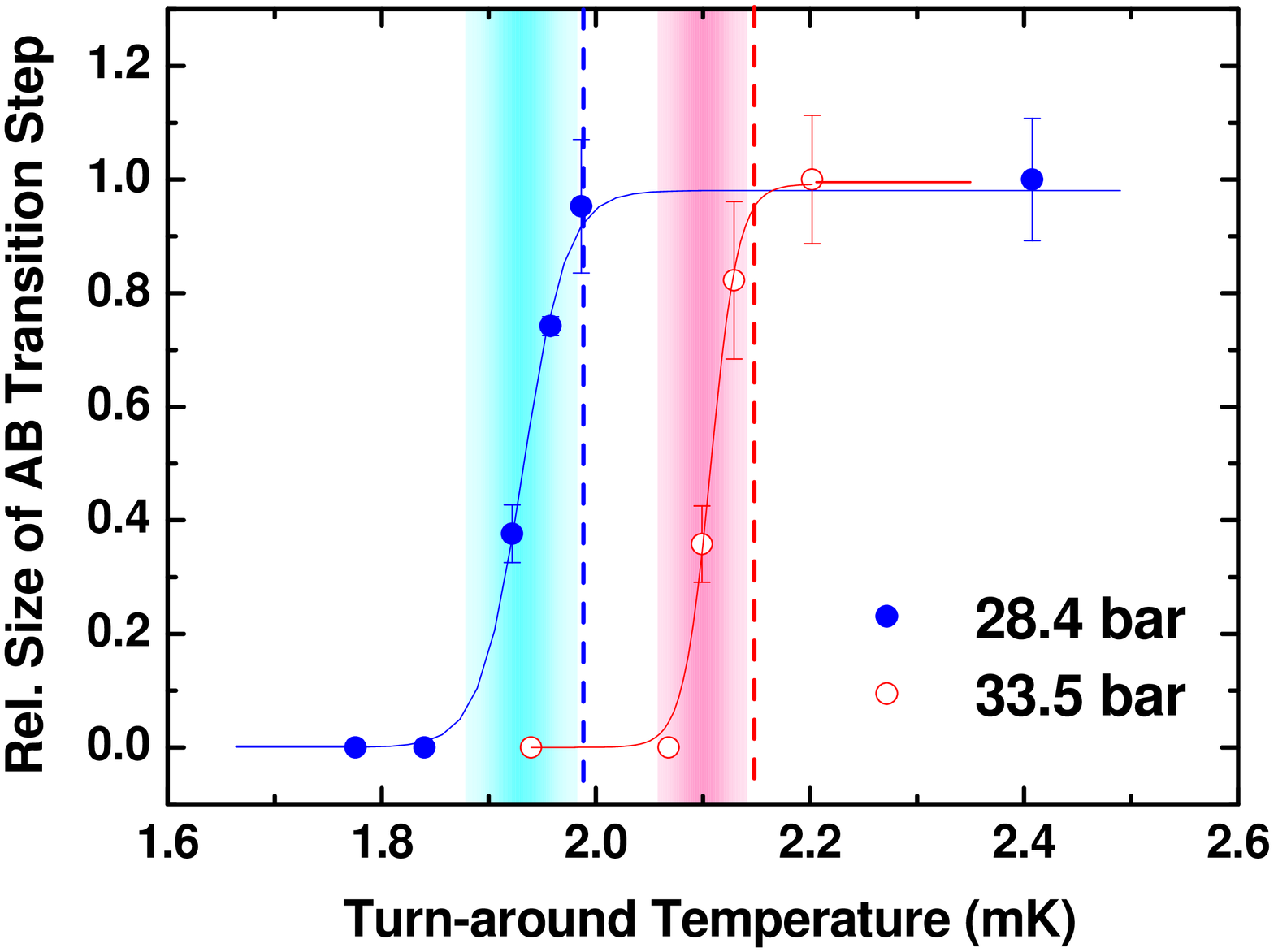}
\caption{\label{jump} The relative size of the steps for the
supercooled aerogel A-B transition is plotted as a function of the
turn-around temperature for 28.4 and 33.5 bar.  The lines going
through the points are guides for eyes.  The dashed vertical lines
indicate the aerogel superfluid transition temperatures.}
\end{figure}

From the data obtained in the tracking experiments, the relative
size of the steps at the supercooled aerogel A-B transition is
plotted in Fig.~2 as a function of the turn-around temperature. For
both pressures we see narrow temperature regions (shaded regions in
the figure) where the size of the steps grows with the turn-around
temperature until $T_{aA}$ is reached.  For $T
> T_{aA}$, no appreciable change in the step size is observed.
This suggests that only a portion of the liquid in aerogel
undergoes the B$\rightarrow$A conversion on warming in that
region.  An inevitable conclusion is that the  A and B phases
coexist in that temperature window.  Consistent behavior has been
observed in 99.3\% porosity aerogel, although in the presence of a
284 G magnetic field \cite{baumgardner}.  It is worthwhile to note
that at 10~G, the equilibrium A-phase width in bulk below the PCP
is $<$ 1~$\mu$K.  A quadratic field dependence in the warming A-B
transition is observed in our study up to 2~kG.  No information on
the spatial distribution of the two phases can be extracted from
our measurements.

In Fig.~3, a composite low temperature phase diagram of $^3$He in
98\% aerogel is reproduced along with our A-B transition
temperatures. We have plotted the lowest temperatures where the
B$\rightarrow$A conversion is first observed on warming. It is
clear that the slope of the A-B transition line in aerogel has the
opposite sign of that in the bulk in the same pressure range.
However, in a weak magnetic field, the slope of the bulk A-B
transition line changes its sign from positive ($p < p_c$ where
$p_c$ represents the polycritical pressure) to negative ($p >
p_c$) (see the dotted line in Fig.~3) \cite{note}. This
observation provides indirect evidence that our measurements take
place below the PCP, if it exists in this system.  It appears that
in effect, the P-T phase diagram is shifted to higher pressures in
the presence of aerogel.

\begin{figure}
\includegraphics[height = 2.3 in]{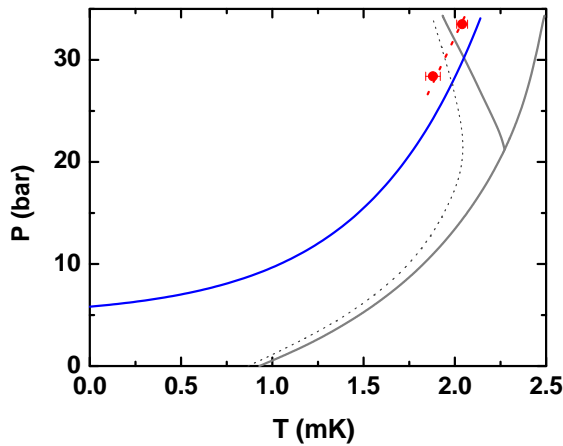}
\caption{\label{phase} The low temperature zero field phase
diagram of superfluid $^3$He in 98.4\% aerogel along with that of
the bulk (grey lines).  The dotted line is the bulk A-B transition
calculated at 1.1~kG from Ref.~\cite{hahn}. The aerogel superfluid
transition line shown in blue (darker line in grey scale) is
obtained by smoothing the results from Ref.~\cite{mat97,gervais}.
Two points are the lowest temperatures where the B$\rightarrow$A
conversion starts on warming.}
\end{figure}

How can we explain the existence of a finite region of the A phase
at pressures below the PCP at $B = 0$? We argue that anisotropic
scattering from the aerogel structure is responsible for this
effect. Although there is no successful quantitative theoretical
account of the A-B transition for $p > p_c$, the Ginzburg-Landau
(GL) theory presents a quantitative picture for the A-B transition
in a small magnetic field, $B$ (relative to the critical field)
for $p < p_c$.  Under these conditions, the quadratic suppression
of the A-B transition arises from a term in the G-L free energy:
\begin{equation}
 f_z = g_{z}B_{\mu}A_{\mu i}A_{\nu i}^*B_{\nu} \label{1}
\end{equation}
where $A_{\mu i}$ represents the order parameter of a superfluid
state with spin ($\mu$) and orbital ($i$) indices
\cite{vollhardt}. The main effect of this term is to produce a
tiny splitting in $T_c$ for the A and B phases.  In the G-L limit,
the free energy (relative to the normal state) of the A (B) phase
is $f_{A(B)} = -\alpha^{2}/{2\beta_{A(B)}}$ where $\alpha =
N(0)(T/T_{c}-1)$ with $N(0)$ being the density of states at the
Fermi surface, and  $\beta_{A(B)}$ is the appropriate combination
of $\beta$-parameters that determine the fourth-order terms in the
G-L theory. For $p < p_{c}$, the B phase has lower free energy
than the A phase ( $\beta_{B} < \beta_{A}$).  However, when the
two phases are highly competing, {\it i.e.} $\beta_{A} \approx
\beta_{B}$, even a tiny splitting in the superfluid transition,
$\delta T_{c} = T_{c}^{A} - T_{c}^{B} ( > 0)$ results in a
substantial temperature region (much larger than $\delta T_c$)
where the A phase becomes stable over the B phase.

The simplified representation of the aerogel as a collection of
homogeneous isotropic scattering centers is not sufficient to
describe minute energy scale phenomena such as the A-B transition.
The strand-like structure introduces an anisotropic nature in the
scattering, {\it e.g.} {\it p-wave} scattering. This consideration
requires an additional term in the G-L free energy
\cite{thunebergb,fomin},
\begin{equation}
 f_a = \alpha_{1}a_{i}A_{\mu i}A_{\mu j}^{*}a_{j} \label{2}
\end{equation}
where $\hat{a}$ is a unit vector pointing in the direction of the
aerogel strand.   In other words, the aerogel strand produces a
random field which couples to the orbital component of the order
parameter. This random orbital field plays a role analogous to the
magnetic field in spin space, thereby splitting the superfluid
transition temperature. If the aerogel structure generates
anisotropy on the length scale of $\xi_o$, $f_{a}$ would give rise
to the A-B transition, even in the absence of a magnetic field.
Detailed free energy considerations indicate that the anisotropy
would favor the $\hat{a} \bot \hat{l}$ configuration for the A
phase \cite{rainer} where $\hat{l}$ indicates the direction of the
nodes in the gap. Using the expression for the coupling strength,
$\alpha_{1}$, calculated in the quasiclassical theory
\cite{thunebergb}, we find that $f_{a}$ is comparable to $f_{z}$
produced by a magnetic field, $B_{e} = \sqrt{\alpha_{1} / g_{z}}
\approx (T_{c} / \gamma\hbar)\sqrt{\xi_{o} / \ell} \sim$ 1 kG
where $\gamma$ is the gyromagnetic ratio of $^3$He and $\ell$ is
the mean free path presented by the impurity scattering off the
aerogel strand.  Since $B_{e} \propto \sqrt{\delta \ell / \ell}$
($f_{a} \propto (\delta \ell / \ell)$), where $\delta \ell / \ell$
represents the anisotropy in the mean free path, only a fraction
of 1\% anisotropy is sufficient to produce the observed A phase
width. The inhomogeneity of the local anisotropy over length
scales larger than $\xi_{o}$ naturally results in the coexistence
of the A and B phase.

A PCP where three phases merge, as in the case of superfluid $^3$He,
should have at least one first order branch \cite{yip}. When this
branch separates two highly competing phases with distinct symmetry,
the PCP is not robust against the presence of disorder. In general,
the coupling of disorder to the distinct order parameters will
produce different free energy contributions for each phase.
Consequently, a strong influence on the PCP is expected under these
circumstances\cite{aoyama}.  It is possible that the PCP vanishes in
response to disorder (as it does in response to a magnetic field)
and a region of coexistence emerges. An experiment on $^3$He-$^4$He
mixtures in high porosity aerogel reported a similar disappearance
of the PCP \cite{kim}. Strikingly similar phenomena have also been
observed in mixed-valent manganites where the structural disorder
introduced by chemical pressure  produces the coexistence of two
highly competing phases (charge ordered and ferromagnetic phases)
separated by a first order transition \cite{dagotto,zhang}. A
growing body of evidence suggests that the coexistence of the two
phases is of fundamental importance in understanding the unusual
colossal magnetoresistance in this material.

Considering the energy scales involved in the A-B transition and
anisotropy, it is not surprising  to see a difference in the
details of the A-B transition in aerogel samples even with the
same porosity.  However, it is important to understand the role of
anisotropic scattering.  We propose that the effect of anisotropic
scattering can be investigated in a systematic manner, at least in
aerogel, by introducing controlled uniaxial stress, which would
generate global anisotropy in addition to the local anisotropy.

\begin{acknowledgments}
This work is partially supported by the NSF through DMR-0239483
(YL) and DMR-0244099 (WPH), and a NHMFL IHRP grant (YL). The
experiment was performed at the High B/T Facility of NHMFL at
University of Florida.  We acknowledge useful discussions with P.
Kumar, A. Biswas, M. Meisel, and J. Sauls. YL thanks the Alfred P.
Sloan Foundation for financial support.
\end{acknowledgments}


\begin{thebibliography}{9}

\bibitem{paLee} P.A. Lee and T.V. Ramakrishnan, Rev. Mod. Phys.
{\bf57}, 287 (1985).
\bibitem{heeger} A.J. Heeger, {\it Solid State Physics} vol. 23, p.283, ed. F. Seitz, D. Turnbull,
and H. Ehrenreich, New York, Academic Press (1969).
\bibitem{sachdev} S. Sachdev, {\it Quantum Phase Transition}, Cambridge
University Press, New York (1999).
\bibitem{vollhardt} D. Vollardt and P. W\"{o}lfle, {\it The
Superfluid Phases of Helium 3}, Taylor \& Francis, London (1990).
\bibitem{porto} J.V. Porto and J.M. Parpia, Phys. Rev. Lett. {\bf
74}, 4667 (1995).
\bibitem{sprague1} D.T. Sprague {\it et al.}, Phys. Rev. Lett. {\bf
75}, 661 (1995).
\bibitem{mat97} K. Matsumoto {\it et al.}, Phys. Rev. Lett. \textbf{79},
253 (1997); Errata ibid. \textbf{79}, 2922 (1997).
\bibitem{gervais} G. Gervais {\it et al.}, Phys. Rev. B {\bf 66},
054528 (2002).
\bibitem{brussard} P. Brussaard {\it et al.}, Phys. Rev. Lett. {\bf
86}, 4580 (2001).
\bibitem{choi} H.C. Choi {\it et al.}, Phys. Rev. Lett. {\bf 93},
145302 (2004).
\bibitem{baumgardner2} J.E. Baumgardner and D.D. Osheroff, Phys.
Rev. Lett. {\bf 93} 155301 (2004).
\bibitem{thuneberg} E.V. Thuneberg {\it et al.} Phys. Rev. Lett. {\bf
80}, 2861 (1998).
\bibitem{barker} B.I. Barker {\it et al.}, Phys. Rev. Lett. \textbf{85}, 2148 (2000).
\bibitem{dmitriev} V.V. Dmitriev {\it et al.}, Physica B {\bf
329}, 320 (2003).
\bibitem{nazaretski} E. Nazaretski, N. Mulders, and J.M. Parpia,
JETP Lett. {\bf 79}, 383 (2004).
\bibitem{baumgardner} J.E. Baumgardner {\it et al.}, Phy. Rev.
Lett. {\bf 93} 055301 (2004).
\bibitem{leggett} A.J. Leggett and S.K. Yip, {\it Helium Three},
p. 523, ed. W.P. Halperin and L.P. Pitaevskii, North-Holland,
Amsterdam (1990).
\bibitem{paulson} D.N. Paulson, H. Kojima, and J.C. Wheatley,
Phys. Rev. Lett. {\bf 32}, 1098 (1974).
\bibitem{li} Y.H. Li and T.L. Ho, Phys. Rev. B {\bf38}, 2362
(1988).
\bibitem{hahn} I. Hahn, Ph.D. thesis, USC (1993).
\bibitem{baramidze} G. Baramidze and G. Kharadze, J. Phys. - Cond.
Mat. {\bf 14}, 7471 (2002).
\bibitem{lee99}Y. Lee {\it et al.}, Nature \textbf{400}, 431 (1999).
\bibitem{note} This behavior persists even in high fields up to the
critical field. The sign cross-over point gradually moves down to
around 19 bar near the critical field.  See Ref. \cite{hahn}.
\bibitem{rainer} D. Rainer and M. Vuorio, J. Phys. C: Solid State
Phys. {\bf 10}, 3093 (1977).
\bibitem{thunebergb} E.V. Thuneberg, cond-mat/9802044 (1998): J.A.
Sauls, private communication.
\bibitem{fomin} I.A. Fomin, JETP {\bf 98}, 974 (2004).
\bibitem{yip} S.K. Yip, T. Li, and P. Kumar, Phys. Rev. B {\bf
43}, 2742 (1991).
\bibitem{aoyama} K. Aoyama and R. Ikeda, cond-mat/0502571v2. This work, which appeared during the review
process of our paper, develops a theoretical analysis of the
influence of local anisotropy on the PCP.
\bibitem{kim} S.B. Kim, J. Ma, and M.H.W. Chan, Phys. Rev. Lett. {\bf
71}, 2268 (1993).
\bibitem{dagotto} E. Dagotto, T. Hotta, and A. Moreo, Phys. Rep.
{\bf 344}, 1 (2001).
\bibitem{zhang} L. Zhang {\it et al.}, Science {\bf 298}, 805
(2002).




\end{thebibliography}
\end{document}